\documentstyle[prd,aps,epsfig,
tighten,
preprint, 
multicol]{revtex}

\begin{document}
\draft


\title{Supersymmetric Neutrino Masses and Mixing with R-parity Violation}

\author{E. J. Chun$^{a}$, S. K. Kang$^{a}$, 
C. W. Kim$^{a,b}$, and U. W. Lee$^{c}$}

\address{ $^{a}$Korea Institute for Advanced Study,
                207-43 Cheongryangri-dong, Dongdaemun-gu,
                Seoul 130-012, Korea}
\address{ $^{b}$Department of Physics \& Astronomy, 
                The Johns Hopkins University,
                Baltimore, MD 21218, USA}
\address{ $^{c}$Department of Physics,
                Mokpo National University,
                Chonnam 534-729, Korea}
\address{Email addresses: 
ejchun@kias.re.kr, skkang@kias.re.kr,
cwkim@kias.re.kr, leeuw@chungkye.mokpo.ac.kr}  

\maketitle

\begin{abstract}
In the context of the minimal supersymmetric standard model,  
nonzero neutrino masses and mixing can be generated through  renormalizable 
lepton number (and thus R-parity) violating operators.
It is examined whether neutrino mass matrices from tree and 
one-loop contributions can account for two mass-squared differences 
and mixing angles that explain current experimental data.
By accommodating, in particular, the solar and atmospheric neutrino data, 
we find interesting restrictions not only on the free parameters
of the theory, such as lepton number violating couplings and soft-parameters,
but also on the oscillation parameters of atmospheric neutrinos.
\end{abstract}

\pacs{PACS number(s): 12.60.Jv, 14.60.Pq}


\maketitle

\section{Introduction}

There exists some evidence  for nonzero neutrino masses and mixing.
The observations of the solar neutrino deficit have been indicating 
neutrino oscillation \cite{BP,SOL}.  
The resonant conversion of $\nu_e$ inside the Sun \cite{MSW} would 
provide the most favorable explanation for the solar neutrino data. 
Several neutrino experiments have also  
observed deficit in the atmospheric neutrino flux \cite{ATM}.
The evidence for atmospheric neutrino oscillations was recently
presented by the Super-Kamiokande group \cite{S-Kam},
which favors  $\nu_\mu$--$\nu_\tau$ (or $\nu_s$) oscillation.  
The $\nu_\mu$--$\nu_e$
oscillation interpretation can be ruled out by CHOOZ experiment \cite{CHOOZ}.
The last one is  laboratory evidence for the neutrino oscillation
$\bar{\nu}_\mu \to \bar{\nu}_e$ coming from the LSND experiment \cite{LSND}.
This evidence is recently being challenged  by KARMEN experiment
\cite{KARMEN}, and is to be checked in the near future.
The above neutrino data are known to require
three distinct mass-squared differences and mixing angles;
\begin{eqnarray}
 \left.
 \begin{array}{rcl}
 \Delta m^2_{\rm sol} &\simeq& (4-10)\times10^{-6}\; {\rm eV}^2 \\
 \sin^22\theta_{\rm sol} &\simeq& (0.12-1.2)\times10^{-2}
 \end{array}  \right\}  && \cite{SOLMSW} \label{SOLMSW} \\
 \left.
 \begin{array}{rcl}
 \Delta m^2_{\rm atm} &\simeq& (0.5-6)\times10^{-3}\; {\rm eV}^2\\
 \sin^22\theta_{\rm atm} &\simeq& (0.82-1)
 \end{array}   \right\}  && \cite{S-Kam,atmrange} \label{ATM} \\
 \left.
 \begin{array}{rcl}
 \Delta m^2_{\rm LSND} &\simeq& (0.3-2.2)\; {\rm eV}^2 \\
 \sin^22\theta_{\rm LSND}  &\simeq& (0.1-4)\times10^{-2}
 \end{array}  \right\} && \cite{LSND} \label{LSND}
\end{eqnarray}
With ordinary three neutrinos,  any two of the mass-squared
differences in the above equations can be obtained: 
that is, those corresponding to (i) solar and atmospheric (S+A),  
(ii) solar and LSND (S+L), or  (iii)  atmospheric and LSND (A+L) 
neutrino data.
For (S+A), the LSND result has to be disregarded.  
In the case of (S+L) or (A+L), the presence of a sterile neutrino 
is necessary for the explanation of the atmospheric 
or solar neutrino experiment, respectively. 

\medskip

One of the desirable features of the supersymmetric extension
of the standard model would be the generation of small neutrino masses 
within its context, as the supersymmetric standard model with 
the minimal particle content (MSSM) allows for the lepton (L) and
baryon number (B) violating operators.
In order to ensure the longevity of a proton, 
one usually assumes the conservation of R-parity, forbidding both 
(renormalizable) B and L violating operators.
As a consequence, the lightest supersymmetric particle (LSP)
is stable and  thus cold dark matter of the universe
may consist of neutral LSP's.
However,  there is no obvious theoretical reason why R-parity needs 
to be conserved, or why both B and L conservation  have to be 
imposed.  L-violation would be present in the MSSM and it may be 
the origin of nonzero neutrino masses and mixing 
that explain current experimental data,
while proton stability is ensured by B conservation alone.

The L-violating operators in the MSSM are 
\begin{equation} \label{Rpv}
 \mu_i L_i H_2\,,\quad  \lambda'_{ijk} L_i Q_j D^c_k\,,\quad 
  {\rm and}  \quad \lambda_{ijk}L_i L_j E^c_k \,. 
\end{equation}
As is well-known \cite{HS}, 
ordinary neutrinos can obtain nonzero masses in tree-level
via nonzero vacuum expectation values (VEVs) of sneutrinos, as well as
at the one-loop level through squark or slepton exchanges.  
Recently, there have been many works studying neutrino phenomenology
in the context of R-parity breaking supersymmetric models \cite{RPN}.
Typically, the tree mass is much larger than the 
loop mass.  The key observation we wish to emphasize is that 
it is, however, possible to find  the soft supersymmetry breaking 
parameter space for which the tree mass 
is rather close to the loop mass, and the solutions to 
the actual neutrino problems can be provided.
This was first recognized by Hempfling in Ref.~\cite{RPN} where 
a scatter plot study of the supersymmetric grand unification 
model allowing only bilinear operators shows that the solar and 
atmospheric neutrino data can be accounted for.  

In this paper, assuming  the presence of trilinear L-violating terms,
we will examine how the soft parameter space viable for neutrino
physics is constrained depending on the choice of the 
L-violating couplings and $\tan\beta$.
Furthermore, we will find that there are certain correlated patterns among
soft parameters and predicted masses and mixing for
given L-violating couplings and $\tan\beta$.

\section{Patterns of Neutrino Mass Matrices}

Before investigating neutrino masses arising in the 
L-violating MSSM, let us first discuss the patterns of mass matrices
required for each case (i), (ii) or (iii).

\begin{center} {\bf (i) (S+A)}\\ \end{center} 
{\it (a) hierarchical neutrino structure}:
The R-parity violation may generate at least two
nonzero mass eigenvalues satisfying $0 < m_{\nu_2} < m_{\nu_3}$, and thus 
$\Delta m^2_{\rm sol} \simeq m^2_{\nu_2}$ and $\Delta m^2_{\rm atm}\simeq
m^2_{\nu_3}$ required for the explanation of the solar and 
atmospheric neutrino data.
Then, we need a mass matrix in the $\nu_\mu,\nu_\tau$ 
directions which yields a large mixing, $\sin^22\theta_{\mu\tau} \sim 1$, 
and a {\it rather small} ratio between two eigenvalues,
\begin{equation}
\chi \equiv m_{\nu_3}/m_{\nu_2} =(7-40)
\end{equation}
as can be obtained from Eqs.(1) and (2).
In this case, the components along the $\nu_e$ direction should be able to
explain the mixing of $\nu_e$ with $\nu_{\mu,\tau}$ 
reproducing $\sin^2{2\theta_{sol}}$ given in Eq.~(1). \\
{\it (b) degenerate neutrino structure}:
Another way to accommodate the solar and atmospheric neutrino data is 
to have almost degenerate three neutrinos.  This could be achieved
in our scheme if the amount of L-violation along the $e,\mu$ and $\tau$ 
directions are the same.  Then, the solar neutrino data could be 
explained by a large mixing resonant conversion or vacuum oscillation effect 
\cite{SOLMSW} which will be disregarded in this paper.

\begin{center} {\bf (ii) (S+L)} \\ \end{center}
For the explanation of the atmospheric neutrino oscillation in addition to
the solar neutrino and LSND data, a sterile neutrino almost maximally 
mixed with $\nu_\mu$ has to be invoked \cite{ckl}.  
Let us denote the $4\times4$
neutrino mass matrix by $m_{ij}$ in $(\nu_e,\nu_\mu,\nu_\tau,\nu_s)$ basis,
\begin{eqnarray}
\left(\begin{array}{cccc}
 m_{ee} & m_{e\mu} & m_{e\tau} & m_{es} \\
 m_{e\mu} & m_{\mu\mu} & m_{\mu \tau} & m_{\mu s} \\
 m_{e\tau} & m_{\mu\tau} & m_{\tau \tau} & m_{\tau s} \\
 m_{es} & m_{\mu s} & m_{\tau s} & m_{ss}  \end{array} \right)
\end{eqnarray}
where the components $m_{i s}$ come from a certain origin beyond the MSSM.
A natural way to obtain a large mixing required by the atmospheric
neutrino experiment is to have the almost Dirac structure with 
$m_{\mu s} \gg m_{es}, m_{\tau s}$.   
There exist two possibilities to realize this (S+L). \\
{\it (a) $m_{\nu_e}, m_{\nu_{\tau}} \gg m_{\nu_{\mu}}, m_{\nu_s}$}:
This is the case where the largest mass scale for $\Delta m^2_{\rm LSND}$ 
is determined by the $(e,\tau)$ block of $m_{ij}$ which is larger than
the $(\mu,s)$ block. As the $\nu_e$--$\nu_\mu$ oscillation explains the LSND
experiment, the solar neutrino data is explained by the $\nu_e$--$\nu_\tau$ 
oscillation.  Therefore, it is required that $\nu_e$ and $\nu_\tau$ are
almost degenerate: $m_{\nu_e} \simeq m_{\nu_\tau} \sim 1$  eV, and 
$m_{\nu_\tau}-m_{\nu_e} \sim 10^{-6} (10^{-10}) $ for the MSW (vacuum)
solution \cite{SOLMSW}.  This extreme degeneracy is, however,
hard to achieve in our scheme as will become clear in Section IV. \\
{\it (b) $ m_{\nu_{\mu}}, m_{\nu_s} \gg  m_{\nu_e}, m_{\nu_{\tau}}$}:
This possibility is to have $\nu_{e, \tau}$ 
lighter than $\nu_{\mu, s}$, for which the solar neutrino data 
can be explained by   small mixing in Eq.~(1).
Then, the mass-squared differences for the atmospheric and LSND data
are determined by 
\begin{equation}
\Delta m^2_{\rm LSND} \simeq m^2_{\mu s}\,, \quad  
\Delta m^2_{\rm atm} \simeq 2m_{\mu s}(m_{\mu\mu}+m_{ss})  \,.
\end{equation}
The mass-squared difference $\Delta  m^2_{\rm atm}$ 
for the $\nu_\mu$--$\nu_s$ oscillation
can be different from that for the $\nu_\mu$--$\nu_\tau$ oscillation
due to matter effects in the Earth.
Recent analysis \cite{FVY} based on the Super-Kamiokande data shows that
$\Delta m^2_{\rm atm}=(0.2-1)\times10^{-2}$ eV$^2$
for the  $\nu_\mu$--$\nu_s$ oscillation
which is  a bit shifted up compared with the value in Eq.~(\ref{ATM}).
For this value of $\Delta m^2_{\rm atm}$ and 
$\Delta m^2_{\rm LSND}$ in Eq.~(3), we get
\begin{eqnarray}
m_{\mu s} &\simeq& (0.55-1.5)~{\rm eV}  \nonumber \\
m_{\mu\mu}+ m_{ss} &\simeq& (1.3-18)\times 10^{-3}~{\rm  eV} \,.
\end{eqnarray}
Since the mixing angle required for the explanation of the LSND result is 
given by $\theta_{\rm LSND} \simeq m_{es}/m_{\mu s}$, one needs
$m_{es} \simeq (2.4-5.5) \times 10^{-2}$ eV.

Let us now estimate the sizes of $m_{ij}$ for active neutrinos
coming from R-parity violation.
If the lepton number violation in the $\mu$ direction is suppressed
so that $m_{\mu\mu},m_{\mu e}, m_{\mu\tau} \ll m_{ss}$, one needs to
generate only the $\nu_e$--$\nu_\tau$ oscillation for the solar neutrino
for suitable values of $m_{e\tau}$ and $m_{\tau\tau}$
which can be rather trivially obtained in terms of the lepton number
violations in the direction of $e$ and $\tau$.   This case requires an
explanation for the origin of the mass scale $m_{ss} \sim 10^{-2}$ eV.
It would be more natural to have $m_{ss} \approx 0$ as in Ref.~\cite{ckl} 
and thus $m_{\mu\mu} \gg m_{ss}$.  In this case, one needs
$m_{\mu\mu} \simeq (1.3-18)\times 10^{-3}$ eV.
Under the assumption that $m_{\tau s}/m_{\mu s} \ll 1$,
the solar neutrino data can be explained if
$m_{\tau\tau}=\sqrt{\Delta m^2_{\rm sol}}$ and
$m_{e\tau}/m_{\tau\tau} - m_{\mu\tau}m_{es}/m_{\tau\tau}m_{\mu s}
=\theta_{\rm sol}$.  {}From this, the component $m_{\mu\tau}$
has to be restricted as $m_{\mu\tau} \lesssim
\sqrt{\Delta m^2_{\rm sol}}\theta_{\rm sol}/\theta_{\rm LSND} $.
Therefore, as in the previous case (i),
the L-violation in the direction of $\mu$ and $\tau$ has to generate
{\it not so large} ratios: 
\begin{equation} \label{caseii}
m_{\mu\mu}/m_{\tau\tau} \simeq (0.41-9)\,, \quad
m_{\mu\tau}/m_{\tau\tau}\lesssim 3.5 \,.  
\end{equation}
For this calculation, we have used the value of $m_{\mu s}$ varying 
from 0.55 eV to 1.5 eV and the largest value of $m_{\mu\tau}$.

\begin{center} {\bf (iii) (A+L)}  \\ \end{center}
This is the most popular case with four neutrinos \cite{FOUR}.  
The solar neutrino problem is solved by $\nu_e$--$\nu_s$ mixing,
and the atmospheric neutrino problem by maximally mixed $\nu_\mu$--$\mu_\tau$
oscillation.  Typically, one would expect that $\nu_{e,s}$ are lighter than
$\nu_{\mu,\tau}$. Combined with the LSND result, it is then required that the 
two mass eigenvalues satisfy 
$m_{\nu_2} \simeq m_{\nu_3} \simeq (0.55-1.5)$ eV 
(therefore $\chi \simeq 1$) and 
$m_{\nu_3}^2-m_{\nu_2}^2 =\Delta m_{\rm atm}^2$.
The LSND mixing angle requires also 
$m_{e\mu}/m_{\mu\mu}$ or $m_{e\tau}/m_{\mu\tau} \simeq 
(2.4-5.5) \times 10^{-2}$.
In the scheme of generating the neutrino masses from the R-parity
violating couplings, it is again difficult to produce a small number:
$(m_{\nu_3}-m_{\nu_2})/(m_{\nu_3}+m_{\nu_2})=\Delta m_{\rm atm}^2/
4m_{\nu_3}^2 \sim 10^{-3}$.
Recall that it is also possible to have almost degenerate $\nu_{e,s}$ with
mass around 1 eV, and lighter $\nu_{\mu,\tau}$.  In this case, one would
need to fine-tune some parameters in the $\nu_e$--$\nu_s$ mass
matrix to achieve  small mixing and a very small mass difference,
$\sim 10^{-6}$ or $10^{-10}$ eV.

\section{MSSM with R-parity and L Violation}

Our framework is the conventional MSSM \cite{SUG} in which 
soft supersymmetry breaking  parameters arise 
from the gravitational coupling to the hidden sector 
and are assumed to be universal at the grand unification (GUT) 
scale where the three gauge couplings meet.
In this framework, there are five independent parameters; 
the scalar mass $m_0$, the trilinear soft parameter $A_0$, 
the gaugino mass $m_{1/2}$, as well as $\tan\beta$ and the sign of $\mu$.   
Here $\mu$ represents the mass parameter of the bilinear operator
$H_1 H_2$.

Without loss of generality, one can take 
only dimensionless L-violating couplings $\lambda'$,
$\lambda$ at the GUT scale.
Recall that, in models where all the L-violating
terms appear generically, the universality condition allows 
us to redefine the superfields $L_i$ and $H_1$ 
in such a way that $\mu_i=D_i=0$ at the GUT scale. 
Here $D_i$ is a dimension-two soft parameter corresponding to
the bilinear operator $L_i H_2$.  
Even if $\mu_i$ and $D_i$ are zero at the GUT scale,
upon renormalizing the soft SUSY breaking  parameters and 
the L-violating couplings down to the weak scale, 
the universality condition breaks down and nonzero $\mu_i$, 
$D_i$ and $m^2_{L_i H_1}$ are generated.  
For the calculation of the  renormalization group equations (RGEs), 
it is convenient to use the basis \cite{CW} where the $\mu_i$ terms are 
continuously rotated away in the following approximate manner:
\begin{eqnarray}
 H_1 + {\mu_i\over \mu} L_i &\longrightarrow& H_1 \\
 L_i - {\mu_i\over \mu} H_1 &\longrightarrow& L_i \nonumber\,.
\end{eqnarray}
This definition of new basis is valid up to the leading order of 
$\lambda'$ and $\lambda$, which can be justified if they are small
enough.  The merit of this basis is that the RGEs of 
$\lambda'$ and $\lambda$ do not mix each other as shown in Appendix.
Nonvanishing $D_i$ and $m^2_{L_i H_1}$ at the weak scale
induce nonzero VEVs of sneutrinos \cite{HS}.  
Keeping only the leading terms of the L-violating soft-parameters, 
one finds the sneutrino VEVs \cite{CW}: 
\begin{equation}  \label{snvev}
 \langle \tilde{\nu}_i\rangle 
 \simeq -{ D_i v_2 + m^2_{L_iH_1} v_1 
 \over m_{L_i}^2 + M_Z^2 \cos2\beta/2} \,.
\end{equation} 
The tree-level neutrino mass matrix is then given by
\begin{equation}  \label{mntree}
 \left( m_\nu^{\rm tree}\right)_{ij} = 
  {\langle \tilde{\nu}_i\rangle \langle \tilde{\nu}_j\rangle 
   \over m_{1/2}/4\pi\alpha_{GUT} + v^2 \sin\beta/\mu} \,.
\end{equation}
Given the L-violating couplings $\lambda'$ and $\lambda$ 
at the weak scale, the neutrinos acquire radiative masses
whose matrix elements are given by
\begin{eqnarray}  \label{mnloop}
 \left( m_\nu^{\rm loop}\right)_{ij} &\simeq& {3 \over 16\pi^2}
    {\Delta_{d_k} m_{d_l}  \over \bar{m}^2_{d_k} }
    \left[ \lambda'_{ikl} \lambda'_{jlk} + (i \leftrightarrow j) \right] 
            \nonumber \\
    &+& {1 \over 16\pi^2}
    {\Delta_{e_k} m_{e_l}  \over \bar{m}^2_{e_k} }
     \left[ \lambda_{ikl} \lambda_{jlk} + (i \leftrightarrow j) \right] \,,
\end{eqnarray}
where $m_{d,e}$ denote the down-type quark and  charged lepton masses,
respectively, $\Delta_{d,e}$ denote the mixing masses of the 
corresponding squarks or sleptons, and $\bar{m}^2$ are the functions of 
the squark or slepton mass eigenvalues $\tilde{m}_{1,2}$
given by $\bar{m}^2= (\tilde{m}_1^2-\tilde{m}_2^2)/
\ln(\tilde{m}_1^2/\tilde{m}_2^2)$ for each generation.  
Here, $i,j,k,l$ are generation indices.
The full neutrino mass matrix is now given by 
$ (m_\nu) = (m_\nu^{\rm tree}) + (m_\nu^{\rm loop})$,
which will be computed with specifying the input values:
the L-violating couplings $\lambda', \lambda$, 
the soft parameters $m_0, A_0, m_{1/2}$ (fixed at the GUT scale), 
as well as $\tan\beta$ and the sign of $\mu$.
For the computation, we took the ranges of the 
soft parameters as follows: $m_0$ from 100 GeV to 1 TeV, 
the absolute value of $A_0$ from 100 GeV to $3 m_0$, and the 
absolute value of $m_{1/2}$ from 100 GeV to 1 TeV, fixing the sign 
of $\mu$ to be positive.   Note that changing the signs of 
$\mu$, $A_0$ and $m_{1/2}$ simultaneously yields the same results.
We also fixed the top quark mass to be 175 GeV,
and  the strong coupling constant $\alpha_s(M_Z)$ to be 0.118.

\medskip

Let us now specify more on the required L-violation.
Since there are too many L-violating couplings, it would be impossible 
to draw sensible conclusions allowing such arbitrariness 
in explaining the neutrino oscillation parameters in Eqs.~(1)--(3).
Our basic assumption in this regard
is to take the couplings with a ``natural" hierarchy: 
the L-violating Yukawa couplings have a hierarchical structure
similar to the corresponding quark or lepton mass matrix.
In other words, e.g., $\lambda'_{ijk} \propto h^d_{jk}$ where 
$h^d_{jk}$ is the Yukawa matrix for the down-type quarks 
before diagonalization. 
This will be a consequence of models that explain the 
fermion hierarchies in terms of flavor symmetry \cite{flavor}.  
In such a scheme, one expects that 
$\lambda'_{i33}$ and  $\lambda_{i33}$ are the largest components 
and  thus give the leading contribution to the neutrino masses.
The subleading contribution comes from the couplings, e.g., 
$\lambda'_{i32}$ or $\lambda'_{i23}$.  Their contribution to the
component $(m_\nu)_{ij}=m_{ij}$ would be suppressed by the factor of 
$(\lambda'_{i32}\lambda'_{j23}/\lambda'_{i33} \lambda'_{j33})
(m_s/m_b)$ compared with the leading contribution.  
It is to be understood that this expression is taken 
after diagonalization of the quark mass matrices.
It was realized recently by Drees et.al. \cite{Drees} that
the ratio ${m_s/m_b}$ could account for the ratio $\chi^{-1}$ 
if $\lambda'_{i32}\lambda'_{j23} \sim \lambda'_{i33}\lambda'_{j33}$.
But, for the couplings with the natural hierarchy, it is expected that
$\lambda'_{i32}\lambda'_{j23}/ \lambda'_{i33}\lambda'_{j33} 
\lesssim m_s/m_b$, and thus the contribution of the 
smaller couplings is suppressed by a factor $(m_s/m_b)^2$.
This is too small to account for the size of various
components of the neutrino mass matrices leading to  
the case (i), (ii) or (iii).  
Therefore, {\it the couplings other than the (33)-components,
$\lambda'_{i33}$ (or $\lambda_{i33}$),  can be neglected} 
in our discussion.
Now, the question one may ask first is 
whether the phenomenologically desirable neutrino mass matrices 
can be obtained by taking a minimal set of the 
L-violating couplings, that is, the trilinear couplings
other than $\lambda'_i\equiv\lambda'_{i33}$ are negligibly small.  
After answering this, we will examine the effect of 
the couplings $\lambda_{i33} \equiv \lambda_i (i \neq 3)$.
We concentrate on the couplings of $\lambda'_{2,3}$ or $\lambda_2$
since the coupling $\lambda'_1$ or $\lambda_1$ (smaller than others) 
can be adjusted to reproduce the desirable mixing of $\nu_e$ 
to other neutrinos as already discussed.

\section{Neutrino Masses from R-parity Violation}

Our task is now to examine whether the neutrino masses coming from
the R-parity violation in the MSSM can reproduce the mass matrix patterns
studied in Section II.
Before presenting our main results, it is useful to look into some
qualitative features of the neutrino masses coming from R-parity violation.
First thing to note is that typically the tree mass is much larger than 
the loop mass. This can be seen by taking a crude approximation to 
integrate the RGEs in the Appendix.
That is, taking the constant coefficients in the RGEs for $D_i$ and 
$m^2_{L_iH_1}$,  one can easily obtain  their values at the weak scale 
which yield the sneutrino VEVs
\begin{equation}
 \langle \tilde{\nu_i}\rangle \approx v_1 {\ln{M_{GUT}\over M_Z} \over 8\pi^2}
 a_i (\lambda'_ih_b + b_i \lambda_i h_\tau)
\end{equation}
where $a_i, b_i$ are the parameters encoding the RGE effects 
for the soft masses and couplings.  Therefore, the tree and loop mass can
be written as
\begin{eqnarray} \label{CRUDE} \nonumber
 \left(m^{\rm tree}_{\nu}\right)_{ij} &\approx& \kappa_0 
          a_ia_j (\lambda'_ih_b + 
 b_i \lambda_i h_\tau)(\lambda'_jh_b + b_j \lambda_j h_\tau) \\
 \left(m^{\rm loop}_{\nu}\right)_{ij} &\approx& \kappa_1
 (\lambda'_i \lambda'_j h_b^2 + b_{ij}\lambda_i\lambda_j h_\tau^2)
\end{eqnarray}
where $\kappa_0 \sim (3\ln(M_{GUT}/M_Z)/8\pi^2)^2 
(M^2_Z \cos^2\beta /m_{1/2}) $ and $\kappa_1\sim 3v_1^2/8\pi^2 \tilde{m}$.  
Generically, one would find
$m^{\rm tree}/m^{\rm loop} \sim \kappa_0/\kappa_1 \gtrsim  10^2$ 
for $\tilde{m}/m_{1/2}\gtrsim 0.1$. The second observation is that, for 
$\lambda'_i \gg \lambda_i$, the tree mass can be aligned with the loop mass;
that is, $a_i \approx a_j$ for $i \neq j$.  The alignment can be weakened
as $\tan\beta$ and $\lambda_i$ become larger.  For a large $\tan\beta$, the 
misalignment of, e.g.,  $A'_3/\lambda'_3-A'_2/\lambda'_2$ can be amplified 
by the RGE effect (and thus $a_3-a_2$ also) since 
\begin{equation}
 8\pi^2 {d\over dt}\!\left({A'_3 \over \lambda'_3} - {A'_2 \over \lambda'_2}
\right) \approx h_\tau A_\tau \,.
\end{equation}
For large $\lambda_i$, it is obvious to have a large misalignment 
since $\lambda_3=0$ due to the antisymmetry of $\lambda_{ijk}$.

\medskip

The above properties play important roles in the case (S+A), for which
one needs {\it two distinctive} nonzero mass eigenvalues.
Since the tree mass is much larger than the loop mass, 
the loop mass can fit better the solar neutrino mass scale with vacuum 
oscillation (requiring $\Delta m^2_{\rm sol} \simeq 10^{-10} {\rm eV}^2$)
while the tree mass gives rise to the atmospheric neutrino mass scale.
This is basically why one finds more soft parameter space for the vacuum 
oscillation solution to the solar neutrino problem than for the MSW solution 
in the scatter plot study by Hempfling \cite{RPN}.  The vacuum oscillation
of solar neutrinos requires a large mixing between $\nu_e$ and $\nu_{\tau}$
in our scheme since the large mixing for vacuum oscillation implies 
$\lambda'_1 \approx \lambda'_{2,3}$.  This might be in conflict with the
Superkamiokande data as well as the CHOOZ result \cite{S-Kam,CHOOZ}.
Therefore, we prefer the small mixing MSW solution to the solar neutrino 
problem as mentioned before.  It is now clear that,
in order to get the neutrino mass hierarchy 
viable for the atmospheric neutrino oscillation and  the MSW solar 
neutrino conversion, the tree mass has to be suppressed requiring
$a_i \ll 1$.  This can occur for certain soft parameters which admit
a cancellation in $\langle \tilde{\nu_i} \rangle$, that is, 
$m^2_{L_i H_1}/D_i \simeq -\tan\beta$.
Then, it could be that the ratio of the tree and loop mass is 
responsible for the two distinctive neutrino masses differing by the factor 
$\chi =(7-40)$.  

\medskip

With the minimal number of L-violating couplings 
(or, $\lambda_i \ll \lambda'_i$), the mere suppression of the tree mass is 
not enough because the tree and loop masses are almost aligned 
as discussed above (that is, $\langle \tilde{\nu}_i \rangle 
\propto \lambda'_i$), which renders the second eigenvalue $m_{\nu_2}$
close to zero.  However, it turns out that 
a partial cancellation between the tree and loop mass can result in 
the breakdown of the alignment and the production of the desirable value 
of $\chi$.  In other words, it is required that 
$|m_\nu^{\rm tree}| \simeq |m_\nu^{\rm loop}|$ and
they have opposite sign.  
In order to quantify these statements, let us give some examples and
calculate the amount of the misalignment measured by 
$\xi \equiv |\lambda'_2 \langle \tilde{\nu}_3\rangle - 
\lambda'_3 \langle \tilde{\nu}_2\rangle|  /
|\lambda'| |\langle \tilde{\nu}\rangle|$ where
$|\lambda'|=\sqrt{{\lambda'_2}^2+{\lambda'_3}^2}$, etc.  
For $\tan\beta=5$, $m_0=201$ GeV, $m_{1/2}=-109$ GeV, and  $A_0=-550$ GeV,
the muon and tau neutrino mass matrix  is found to be
\begin{equation} \label{E1}
m_\nu /\mbox{eV}\approx  -\pmatrix{ 0.67 & 0.68 \cr 0.68 & 0.69 \cr} + 
                 \pmatrix{ 0.65 & 0.65 \cr 0.65 & 0.65 \cr}
\end{equation}
where the first (second) one is tree (loop) contribution.
As is obvious from Eq.~(\ref{E1}), the tree and loop masses are
aligned very closely ($\xi=7\times10^{-3}$)
and the cancellation of the largest digit results in 
$\chi \approx 38$ and $\sin^22\theta \approx 0.9$.  
For $\tan\beta=40$, taking $m_0=20$ GeV, $m_{1/2}=-110$ GeV, $A_0=-250$ GeV,
one finds  the mass matrix with $\chi \approx 16$ and 
$\sin^22\theta \approx 0.9$;
\begin{equation}
m_\nu /\mbox{eV}\approx  -\pmatrix{ 0.69 & 1.22 \cr 1.22 & 2.14 \cr} + 
                 \pmatrix{ 4.01 & 3.87 \cr 3.87 & 3.73 \cr}
\end{equation}
which shows that  the alignment is badly broken ($\xi=0.1$). 
Still, one needs a cancellation
to get the right value for the mass ratio, in particular.
It turns out that negative values of $m_{1/2}$ and $A_0$ are necessary
to yield opposite signs for the tree and loop masses.
Now that the alignment becomes severer for smaller $\tan\beta$,
the soft parameters have to be tuned more precisely for smaller $\tan\beta$.  
In the numerical computation, 
we calculated the ratio $\chi$ of two mass eigenvalues and the
atmospheric neutrino mixing $\sin^22\theta$ for some sample set of 
$\tan\beta$ and the ratios between the couplings $\lambda'_{2,3}$,
scanning the soft parameters up to the smallest digit above point.
The results of the computations are presented in TABLE I and II.
Notice that the neutrino mass matrix is proportional to 
the overall scale of ${\lambda'}^2_{2,3}$, and thus 
the ratio $\chi$ and the mixing $\sin^22\theta$
depend only on the ratio of two couplings $\lambda'_3/\lambda'_2$.
As expected, the number of the desirable soft parameters 
gets smaller for smaller $\tan\beta$, and 
we find {\it no parameter space for} $\tan\beta \lesssim 4$.
Furthermore, the values of $\chi$ and $\sin^22\theta$  depend 
very sensitively on the soft parameters, and the acceptable soft 
parameters are scattered for a small $\tan\beta$.
For a large $\tan\beta$, $\chi$ becomes a slowly varying function of the
soft parameters, and one can isolate some finite region for given
values of $\chi$ and $\sin^22\theta$.

Unexpectedly, some patterns have emerged for 
values of the ratio $\chi$ and the mixing angle that are  
realized in our scheme depending on $\tan\beta$ and the L-violating  
couplings, as presented in TABLE I.  
For illustration, we took some variations of 
$\lambda'_3$ larger than $\lambda'_2$. 
Similar results can be obtained also in the cases with $\lambda'_2$ 
larger than $\lambda'_3$.
For given $\tan\beta$ and $\lambda'_{2,3}$, 
$\chi$ and $\sin^22\theta$ are found 
to lie inside some restricted ranges, and to be correlated 
in the way that {\it larger mass ratio corresponds to larger mixing
for $\lambda'_2=\lambda'_3$,
or to  smaller mixing for $\lambda'_2 < \lambda'_3$}.
The correlation becomes weaker for larger $\tan\beta$.
{}From TABLE I,  one sees that the maximal mixing is easily realized 
with $\lambda'_3/\lambda'_2 \simeq 2$.
We found practically {\it no desirable parameter space for }
$\lambda'_3/\lambda'_2 \gtrsim 5$ which appears inconsistent with
large $\sin^22\theta$.
The eigenvalue $m_{\nu_2}$ depends on $m_0$ in the way that 
larger $m_0$ produces smaller neutrino mass as implied by 
the formulae (\ref{snvev})--(\ref{mnloop}).  
The right values of the L-violating couplings can be obtained by
the rescaling: 
$\lambda'_{2,3}=10^{-4}(\Delta m^2_{\rm sol}/m_{\nu_2}^2)$ where
$m_{\nu_2}$ is a value given in TABLE I.  For instances, 
the actual values one needs are 
$\lambda'_2=\lambda'_3= (10^{-3}-10^{-4})$ for $\tan\beta=5$, 
$\lambda'_2=\lambda'_3= (10^{-4}-10^{-5})$ for $\tan\beta=20$ and 
$\lambda'_2=\lambda'_3= (3\times10^{-5}-3\times10^{-6})$ 
for $\tan\beta=40$.
As a consequence of  partial cancellation between the tree
and loop mass, the size of the L-violating couplings  
yielding the phenomenologically viable neutrino masses becomes slightly
larger than the values for which the loop mass yields the 
atmospheric neutrino mass scale, say $5\times 10^{-2}$ eV.

It is also amusing to find that 
there are  preferable ranges of the acceptable soft parameters: 
{\it  larger $|A_0|$ and  smaller $|m_{1/2}|$} 
are preferred.   To be specific, we show in TABLE II 
the ranges of soft parameters within which viable neutrino masses 
and mixing can be realized with $\lambda'_2=\lambda'_3=10^{-4}$.
For a given $m_0$, the values of $A_0$ and $m_{1/2}$ should reside
in between two end values shown in the table.  
There is a correlation between $A_0$ and $m_{1/2}$:
larger $|A_0|$ corresponds to larger $|m_{1/2}|$.
A similar pattern follows even with a slight variation of two couplings
$\lambda'_{2,3}$ such as given in TABLE I.

\medskip

If the coupling $\lambda_2$ is comparable to $\lambda'_{2,3}$, then 
the above mentioned properties are significantly modified as anticipated
in the beginning of this section. 
We find that $\chi$ and $\sin^22\theta$ become very slowly varying 
functions of the soft parameters even for  a small $\tan\beta$, 
as a consequence of a large misalignment between $\lambda'_i$ and 
$\langle \tilde{\nu}_i \rangle$.
But one still needs $m^{\rm tree}_\nu\sim m^{\rm loop}_\nu$ 
requiring a cancellation in the sneutrino VEVs, which essentially 
restricts the soft parameter space.  To show this explicitly, let us
take two examples for $\tan\beta=5$.  
Taking $m_0=500$ GeV, $m_{1/2}=-100$ GeV and $A_0=-780$ GeV, one finds 
the mass matrix in which the tree mass is larger than the loop mass as 
in the above minimal cases:
\begin{equation}
m_\nu /\mbox{eV}\approx -\pmatrix{ 0.581 & 0.527 \cr 0.527 & 0.479 \cr}
              +\pmatrix{ 0.394 & 0.390 \cr 0.390 & 0.390 \cr}
\end{equation}
giving $\chi\approx 38$ and $\sin^22\theta\approx 0.89$.
There is a small contribution (0.003 eV) to the 
$m^{\rm loop}_{\nu_\mu \nu_\mu}$ 
from the $\lambda_2$ coupling.  The other one is of different type 
with dominant loop contribution: for $m_0=500$ GeV, $m_{1/2}=400$ GeV and 
$A_0=100$ GeV, one gets 
\begin{equation}
m_\nu /\mbox{eV}\approx -\pmatrix{ 0.071 & 0.023\cr 0.023& 0.0077 \cr}
              +\pmatrix{ 0.136 & 0.127 \cr 0.127 & 0.127 \cr}
\end{equation}
which gives $\chi\approx 20$ and  $\sin^22\theta\approx 0.95$. 
In this case, the $\lambda_2$ contribution (0.009 eV) is more significant 
than before.  In both examples, any of the tree and loop contributions
cannot be neglected in order to produce the right values of $\chi$ and 
$\sin^22\theta$.
%
In TABLE III, we illustrate some set of soft parameter ranges yielding 
the right values of the ratio $\chi$ and the mixing angle,
taking $\lambda'_2=\lambda'_3=\lambda_2$.  
The ranges of $A_0$ and $m_{1/2}$ for  given $m_0$ are shown.  
In TABLE IV, we take some variations of $\lambda_2/\lambda'_{2,3}$.
Compared to the previous case, one finds more parameter space open up  
for larger $\lambda_2$.  To have the right value of $m_{\nu_2}$, we need
a bit smaller values for the couplings $\lambda'$ than before: 
that is, 
$\lambda'_{2,3}=\lambda_2 \simeq (10^{-4} -3\times10^{-5})$ 
for $\tan\beta=5$ and 
$\lambda'_{2,3}=\lambda_2 \simeq (10^{-5} -2\times10^{-6})$ 
for $\tan\beta=40$. 
{}From TABLE III and IV, one can  also see that larger $\tan\beta$ 
are still needed  to destroy the alignment in a sufficient amount 
even with sizable $\lambda_2$.  It can be found that a large soft parameter 
space opens up for $\tan\beta \lesssim2$ only when 
$\lambda_2\gtrsim 4 \lambda'_2$.
Contrary to the case with negligible $\lambda_2$, almost all ranges of
$\chi$, $\sin^22\theta$ can be realized as shown in TABLE III.

\bigskip 

Now let us turn to the other cases.
For the case (S+L),  it is important to realize that
one {\it  needs not to generate two nonzero eigenvalues} 
in the neutrino mass matrix along the $\nu_\mu, \nu_\tau$ directions.  
Furthermore, the ratios $m_{\mu\tau}/m_{\tau\tau}$ and   
$m_{\mu\mu}/m_{\tau\tau}$ (\ref{caseii}) are determined roughly 
by the input values $\lambda'_2/\lambda'_3$ and 
$(\lambda'_2/\lambda'_3)^2$, respectively.
Therefore, it is required from the previous discussion,
$\lambda_2'/\lambda_3' \simeq (0.64-3)$, and the value of
$\lambda'_3$ to be determined by the condition 
$m^\nu_{\tau\tau}= \sqrt{ \Delta m^2_{\rm sol}} \sim 10^{-3} {\rm eV}$.
Generically, the tree mass is much larger than the loop mass.
Taking $\lambda'_{i33} = 10^{-4}$, the tree mass can be 
as large as 100 eV for a small $\tan\beta$ \cite{RPN}.
Therefore, for a small $\tan\beta$,
one needs $\lambda'_3 \gtrsim 10^{-7}$.  This value can be
as large as $10^{-4}$ when the tree mass is suppressed
as we discuss above.  Since $\lambda_i$ cannot give rise to
the nonzero component $m^\nu_{\tau\tau}$, taking the largest 
components being $\lambda'_2 \sim \lambda'_3$ as above is the
best way to explain
$m^\nu_{\mu\mu} \sim m^\nu_{\mu\tau} \sim m^\nu_{\tau\tau}$.

Let us finally comment on the case (A+L).
In order to achieve $(m_{\nu_3}-m_{\nu_2})/(m_{\nu_3}+m_{\nu_2})
\sim 10^{-3}$, much finer tuning of the soft parameters is 
required and thus it is very difficult to be realized in our scheme.

\section{Conclusions}

In conclusion,  we have examined the possibility of obtaining the
realistic neutrino masses and mixing in the context of the
R-parity violating minimal supersymmetric standard model.
We have assumed an ultraviolet theory which has generic L-violating
Yukawa couplings in the basis where the L-violating bilinear
terms are rotated away.  The L-violating Yukawa couplings
induce the L-violating bilinear soft terms through the 
renormalization group evolution which takes a simple form 
in the basis (valid for  small L-violating couplings) as shown
in Appendix.  Analyzing the neutrino masses arising both from the
sneutrino vacuum expectation values generated by the L-violating
soft terms and from the loop corrections through squark or slepton
exchange, we found restrictions on the soft parameters, the
L-violating couplings and $\tan\beta$, under which realistic
neutrino mass matrices can be obtained.

\medskip

With three ordinary neutrinos, one can account for
any two of the three distinct mass-squared differences 
required by the solar, atmospheric and LSND neutrino data.
First, we have discussed the phenomenological mass matrices 
along the $\nu_\mu, \nu_\tau$ directions  which are required by
the data.
If the solar neutrino and LSND data are due to the active neutrinos, and 
a sterile neutrino is introduced to explain the atmospheric data,
then it is not necessary to have two distinct mass eigenvalues
for the mass matrix in the $\nu_\mu,\nu_\tau$ directions but its
components should not differ by  a factor of more than a few.
This can be easily achieved if $\lambda'_{i33}$ are the largest 
L-violating couplings and $\lambda'_{233}\sim \lambda'_{333}$.
In the case of solving the atmospheric neutrino and LSND data, 
one needs to generate two almost degenerate mass eigenvalues
for $\nu_\mu,\nu_\tau$ with a very small mass difference.
This case can hardly be realizable in our scheme.

\medskip

Most nontrivial and realistic  case is to accommodate the
solar and atmospheric data within the ordinary context of
three active neutrinos (disregarding the LSND data).
In this case, one needs to generate two distinct mass eigenvalues
for $\nu_\mu,\nu_\tau$ whose ratio, $\chi$,
should reside roughly between 7 and 40, and 
the mixing angle $\sin^22\theta_{\rm atm} \gtrsim 0.82$.

Under the assumption of the natural Yukawa hierarchy in the 
L-violating couplings, we have argued that the relevant contribution
to the phenomenological neutrino mass matrices 
comes from the components $\lambda'_{i33}$ and $\lambda_{i33}$.
With the minimal choice of the L-violating couplings (namely,
other than $\lambda'_{i33}$ are negligible),  
we needed not only the suppression of tree mass 
(that is, $\tan\beta \simeq -m^2_{L_i H_1}/D_i$), but also 
some partial cancellation between the tree and loop mass.
This basically strongly constrains the soft parameter space.
Our study have shown that the realistic neutrino masses and mixing 
prefer a large trilinear soft parameter $A_0$ and a  small 
gaugino mass $m_{1/2}$.  The desirable soft parameter space becomes
more restricted and finer tuned for smaller $\tan\beta$, 
and thus a reasonable parameter space can be found
only for fairly large $\tan\beta$, say, $\tan\beta \gtrsim 40$.
We have found indeed no parameter space for $\tan\beta \lesssim 4$.
The large mixing explaining the atmospheric neutrino data requires 
that $\lambda'_{233}$ and $\lambda'_{333}$ should not 
differ by more than a factor of 5.
Interestingly, the mass ratio $\chi=m_{\nu_3}/m_{\nu_2}$
and the atmospheric neutrino  mixing $\sin^22\theta$ are 
found to be restricted in a certain range and correlated in the way that 
smaller ratio $\chi$ has smaller or larger  mixing depending on the values
of $\lambda'_{2,3}$.  This tendency becomes also weaker for larger $\tan\beta$
and larger $\lambda_{233}$.
%
It appears more reasonable to have a sizable $\lambda_{233}$, for which 
the experimental quantities of neutrino oscillations become very
stable under the variation of the soft parameters, and thus there exist 
fairly extended regions of parameters fitting into the experimental data. 
Still, there exist significant constraints on the soft parameters
coming from the fact that the tree mass has to be suppressed as above.
More soft parameter space can be found  for larger $\tan\beta$ and 
$\lambda_{233}$.

\section*{APPENDIX}

Renormalization group equations for the lepton number violating 
parameters in the basis where $L_iH_1$ terms are rotated away in the
superpotential.
\begin{eqnarray}
16\pi^2 {d \lambda'_i \over d t} &=& 
\lambda'_i ( \delta_{i3} h_\tau^2 + h_t^2 + 3 h_b^2 
  - {7\over9} g_1^2 - 3 g_2^2 -{16\over3}g_3^2) \\
16\pi^2 {d \lambda_i \over d t} &=& 
\lambda_i ( 3 h_\tau^2  - 3 g_1^2 - 3 g_2^2) \\
16\pi^2 {d A'_i \over d t} &=& 
A'_i ( \delta_{i3} h_\tau^2 + h_t^2 + 9 h_b^2 
      - {7\over9} g_1^2 - 3 g_2^2 -{16\over3}g_3^2) 
+ A_i(2h_b h_\tau) \\
&+ & 2 \lambda'_i ( \delta_{i3} h_\tau A_\tau + h_t A_t + 2 h_b A_b
            + {7\over9} g_1^2 M_1+ 3 g_2^2M_2 +{16\over3}g_3^2M_3)
             \nonumber\\
16\pi^2 {d A_i \over d t} &=& 
A_i ( 5 h_\tau^2  - 3g_1^2 - 3 g_2^2)
+ A'_i(6h_b h_\tau) 
+\lambda_i ( 6 h_\tau A_\tau + 6 g_1^2 M_1+ 6 g_2^2M_2)  \\
16\pi^2 {d m_{L_i H_1}^2 \over d t} &=& 
m_{L_i H_1}^2 ( 3 \delta_{i3} h_\tau^2 + h_\tau^2 + 3 h_b^2 ) 
-6A'_i A_b - 2 A_i A_\tau  \\
&-&6\lambda'_i h_b (m_{L_i}^2+m_{Q_3}^2+m_{D_3}^2) 
-2\lambda_i h_\tau (m_{L_i}^2+m_{L_3}^2+m_{E_3}^2) \nonumber \\
16\pi^2 {d D_i \over d t} &=& 
D_i (3 h_t^2-g_1^2-3g_2^2) -\mu (6 A'_i h_b+2 A_i h_\tau)
\end{eqnarray}

\acknowledgments  
EJC would like to thank B.~de Carlos for useful communications.
UWL thanks KIAS for the kind hospitality during his visit, and BSRI
of Mokpo National University.
EJC is supported by Non-Directed Research Fund of 
Korea Research Foundation, 1996.


\begin{table}
\begin{center}
\caption{Ranges of the atmospheric neutrino mixing angle, the ratio
of two mass eigenvalues corresponding to $\sqrt{\Delta m^2_{\rm atm}/
\Delta m^2_{\rm sol}}$, and  the smallest mass eigenvalue (for solar neutrino)
which can be realized in the soft parameter space
for given $\tan\beta$ and L-violating couplings $\lambda'_{2,3}$. 
$\lambda'_2=10^{-4}$ is taken.
}

\begin{tabular}{ccccc}
 &$\tan\beta$          &    5   &    20 &     40 \\  \hline
 &$\sin^22\theta$      & $0.82\sim0.91$ & $0.82\sim0.95$ & $0.82\sim1.0$  \\
 $\lambda'_3/\lambda'_2=1$ 
 &$m_{\nu_3}/m_{\nu_2}$ & $19\sim40$    &  $14\sim40$  & $7\sim40$   \\
 &$m_{\nu_2}$ (eV)      & $3\times10^{-5}-3\times10^{-3}$ &
                        $10^{-3}\sim0.1$ &  $0.03\sim2$ \\ \hline
 &$\sin^22\theta$    & $0.89\sim1.0$ & $0.86\sim1.0$ & $0.82\sim1.0$ \\
 $\lambda'_3/\lambda'_2=2$ 
 &$m_{\nu_3}/m_{\nu_2}$ & $40\sim7$  & $40\sim7$  & $7\sim40$ \\
 &$m_{\nu_2}$ (eV)      & $3\times10^{-3}\sim10^{-2}$ &   
                        $0.01\sim0.4$ &  $0.05\sim4$ \\  \hline
 &$\sin^22\theta$      & $0.82\sim0.96$ & $0.82\sim0.97$ & $0.82\sim0.99$ \\
 $\lambda'_3/\lambda'_2=3$ 
 &$m_{\nu_3}/m_{\nu_2}$ & $16\sim7$   &  $16\sim7$   & $28 \sim 7$  \\
 &$m_{\nu_2}$ (eV)      & $2\times10^{-4}\sim2\times10^{-2}$ &
                        $5\times10^{-3}\sim0.8$ &  $0.1\sim5$ \\ \hline
 &$\sin^22\theta$      & $0.82\sim0.88$ & $0.82\sim0.89$ & $0.82\sim0.94$  \\ 
 $\lambda'_3/\lambda'_2=4$ 
 &$m_{\nu_3}/m_{\nu_2}$ & $9\sim7$ &  $9\sim7$ & $9\sim7$  \\ 
 &$m_{\nu_2}$ (eV)      & $4\times10^{-4}\sim2\times10^{-2}$ &     
                        $8\times10^{-3}\sim1$ &  $0.4\sim2$ \\ 
\end{tabular}

\bigskip

\caption{ Illustrated ranges of soft parameters  in units of GeV 
within which the solar and atmospheric neutrino data can be accommodated
by suitable choices of soft parameters in the case of 
$\lambda'_{2}=\lambda'_3=10^{-4}$.
The sign of $\mu$ is fixed to be positive.  }

\begin{tabular}{cccc}
 $\tan\beta$ & $m_0$ & $A_0$ & $m_{1/2}$ \\ \hline
 5 & 200 &  $-590 \sim -500$  &  $-150 \sim -100$ \\
   & 400 &  $-1000 \sim -730$ &  $-270 \sim -100$ \\
   & 600 &  $-950 \sim -900$ &  $-150 \sim -100$ \\ \hline
20 & 200 &  $-600 \sim -230$ &  $-270 \sim-100$  \\
   & 500 &  $-1000 \sim-300$  &  $-430 \sim -150$ \\
   & 800 &  $-1000 \sim -370$ &  $-430 \sim -100$ \\ \hline
40 & 200 &  $-600 \sim -200$ &  $-300 \sim -100$ \\
   & 500 &  $-1000 \sim -500$ &  $-500 \sim -100$ \\
   & 800 &  $-1000\sim -240$ &  $-480\sim -100$ \\
\end{tabular}

\bigskip

\caption{Illustrated set of soft parameter ranges  yielding realistic values
of the mixing and the mass ratio in the case of 
$\lambda'_2=\lambda'_3=\lambda_2=10^{-4}$.
The sign of $\mu$ is again taken to be positive. 
Soft parameters are in units of GeV and 
neutrino masses are in units of eV.}

\begin{tabular}{ccccccc}
$\tan\beta$ & $m_0$ & $A_0$ & $m_{1/2}$  & $m_{\nu_2}$  &
  $\sin^22\theta$  & $m_{\nu_3}/m_{\nu_2}$  \\  \hline
 &100 & $-100 \sim -170$ & 430 $\sim$ 570  & $(2 \sim 7)\times10^{-2}$ &
     0.82 $\sim$ 0.96 & 7 $\sim$ 40  \\
3& 300 & $-310\sim -830$ & 490 $\sim$ 210 & $(1 \sim 3)\times10^{-2}$ &
     0.82 $\sim$ 0.94 & 7 $\sim$ 40 \\
 & 500 & $-830$ $\sim$ $-970$ & 510 $\sim$ 470 & $(0.9\sim1.1)\times10^{-2}$ &
     0.82 $\sim$ 0.88 & 25 $\sim$ 40  \\ \hline

 & 100 & 100 $\sim$ 310  & 160 $\sim$ 490  & $(5 \sim 16)\times10^{-2}$ &
      0.82 $\sim$ 0.91  & 7 $\sim$ 40 \\
 & 500 & $-770$ $\sim$ $-990$ & $-100$ $\sim$ $-240$ &
       $(0.6 \sim 2)\times10^{-2}$ & 0.82 $\sim$  1.0 & 7 $\sim$ 40 \\
5&     & 100 $\sim$ 530  & 370 $\sim$ 1000 & $(1 \sim 3)\times10^{-2}$ &
        0.82 $\sim$ 1.0 & 7 $\sim$ 40 \\
 & 900 & $-100 \sim -170$ & 690 $\sim$ 550 & $(5 \sim 6)\times10^{-3}$ &
       0.99 $\sim$ 1.0 & 37 $\sim$ 40 \\
 &     & 100 $\sim$ 530 & 690 $\sim$ 1000 & $(5 \sim 10)\times10^{-3}$ &
       0.88 $\sim$ 1.0 & 14 $\sim$ 40  \\ \hline

 & 300 & $-630$ $\sim$ $-910$ & $-290$ $\sim$ $-430$ & 2.8 $\sim$ 3.5  &
       0.82 $\sim$ 1.0 & 7 $\sim$ 12 \\
 &     & 530 $\sim$ 910 & 290 $\sim$ 510 & 2.5 $\sim$ 3.5  &
       0.82 $\sim$ 1.0  & 7 $\sim$ 14  \\
40& 500 & $-250$ $\sim$ $-330$ & $-100$ $\sim$ $-130$ & 0.4 $\sim$ 0.9  &
       0.82 $\sim$ 1.0  &  7 $\sim$ 19  \\
 &     & 930 $\sim$ 1000  & 510 $\sim$ 570  & 1 $\sim$ 2   &
       0.82 $\sim$ 1.0 & 7 $\sim$ 9 \\
 & 900 & $-300$ $\sim$ $-590$ & $-100$ $\sim$ $-230$ & 0.1 $\sim$ 0.5   &
       0.82 $\sim$ 1.0 & 7 $\sim$ 40  \\
 &     & 100 $\sim$ 280 & 110 $\sim$ 210 & 0.2 $\sim$ 0.6 &
       0.82 $\sim$ 1.0  & 7 $\sim$ 40
\end{tabular}

\bigskip

\caption{Same as in TABLE III with different set of $\lambda_2$. Here
two approximate end values of soft parameters $A_0$ and $m_{1/2}$ 
with the corresponding neutrino parameters are shown.
}

\begin{tabular}{cc|cc|cc}
&& $\lambda_2=3\times10^{-5}$ && $\lambda_2=3\times10^{-4}$ &  \\ \hline
$\tan\beta$ & $m_0$ & $A_0$ & $m_{1/2}$   
                    & $A_0$ & $m_{1/2}$   \\  \hline

3&100 & $-300\sim-120$ & $120\sim240$  
      & $100\sim160$ & $380\sim1000$  \\
 &300 & $-900\sim-640 $ & $320\sim420$  
      & $-100\sim-900 $ & $150\sim1000$   \\
 &900 &      none       &  none              
      & $-1000\sim-620 $ & $880\sim1000$   \\ \hline

5&100 & $100\sim300 $ & $140\sim380$  
      & $100\sim300$  & $220\sim1000$  \\
 &500 & $-1000\sim-820$ & $-220\sim-120$ 
      & $-1000\sim-100$  & $-200\sim 460$ \\
 &    &  none  & none  
      & $100\sim340$  & $580\sim 1000$ \\
 &900 & none & none 
      & $-1000\sim-100$ & $120\sim1000$  \\ \hline

40&300 & $-300\sim-240$ & $-120\sim-100$ 
       & $-600\sim-320$    & $-320\sim-160$ \\
  &    & $180\sim260$ & $120\sim140$ 
       & $240\sim560$   & $160\sim360$ \\
 &500 & $-560\sim-300$ & $-240\sim-120$ 
      & $-1000\sim-540$ & $-540\sim-260$  \\
 &    & $140\sim460$ & $120\sim240$ 
      & $400\sim980$ & $280\sim680$  \\
 &900 & $-1000\sim-420$ & $-420\sim-100$ 
      & $-320\sim-300$ & $-100\sim-120$ \\
 &    & $100\sim800$ & $120\sim480$ 
      & $740\sim1000$ & $520\sim680$  \\
\end{tabular}

\end{center}
\end{table}

\end{document}